\documentclass[11pt]{article}
 \usepackage{amsfonts}

 \newcommand{\sect}[1]{\setcounter{equation}{0}\section{#1}}

 \def\be{\begin{equation}}
 \def\ee{\end{equation}}
 \def\bea{\begin{eqnarray}}
 \def\eea{\end{eqnarray}}

 \def\1{\'{\i}}                           
 \def\R{{\mathbb R}} 
 \def\>#1{{\bf #1}}                 
 
 \def\d{{\rm d}}

 \def\te{\theta}
 \def\rr{\rho}

 \def\jp{J_+}
 \def\jm{J_-}
 \def\jj{J_3}

 \def\otra{b}
 \def\co{\Delta}

 \def\SW{\rm I}
 \def\Stc{\rm S}

 \def\la{\lambda}

 \def\otra{b}

\begin{document}

 \thispagestyle{empty}

 \ 
 \hfill\

 \begin{center}

 {\LARGE{\bf{Curvature from quantum deformations}}}

 \bigskip

 \bigskip

 \end{center}

 \bigskip 
 \medskip

 \begin{center}   {\sc Angel~Ballesteros$^a$,
  Francisco~J.~Herranz$^a$ and Orlando Ragnisco$^{b}$ }
 \end{center}

 \begin{center} {\it { 
 ${}^a$Departamento de F\1sica, Universidad de Burgos, Pza.\
 Misael Ba\~nuelos s.n., \\ E-09001 Burgos, Spain }}\\ e-mail:
 angelb@ubu.es, fjherranz@ubu.es
 \end{center}

 \begin{center} {\it { 
 ${}^b$Dipartimento di Fisica,   Universit\`a di Roma Tre and
 Instituto Nazionale di Fisica Nucleare sezione di Roma Tre,  Via Vasca
 Navale 84,\\ I-00146 Roma, Italy}}\\ e-mail: ragnisco@fis.uniroma3.it
 \end{center}

 \bigskip 
 \medskip

 \begin{abstract}

 A Poisson coalgebra analogue of a (non-standard) quantum deformation
 of $sl(2)$ is shown to generate an integrable geodesic dynamics on
 certain 2D spaces of non-constant curvature. Such a curvature   
 depends on the quantum deformation parameter $z$ and the flat case is
 recovered in the limit
 ${z\to 0}$. A superintegrable geodesic dynamics can also be  defined
 in the same framework, and the corresponding spaces turn out to be
 either   Riemannian or relativistic spacetimes (AdS and dS)   with
 constant curvature equal to $z$.  The underlying coalgebra symmetry of
 this approach ensures the existence of its generalization to arbitrary
 dimension.
 \end{abstract}

 \bigskip\medskip

 \bigskip\medskip 



 \newpage


 \sect{Introduction}

 Both theory and applications of quantum groups
 have certainly motivated an intensive effort  aimed at understanding
 the role played by Hopf algebra deformations from many different
 viewpoints (see, for instance,~\cite{Dri,CP,majid,Sierra}). In
 particular, Poisson Hopf algebras (namely, Poisson--Lie groups and
 their associated Lie bialgebras and Drinfel'd doubles~\cite{Dr}) are
 just the Poisson counterparts of quantum groups and algebras.
 Recently, a systematic approach to the construction of integrable and
 superintegrable Hamiltonian systems from Poisson coalgebras has been
 introduced (see~\cite{BR,Deform,cluster} and references therein). In
 this context, Poisson coalgebras associated to quantum groups  can be
 understood as the dynamical symmetries that generate integrable
 deformations of well-known dynamical systems with an arbitrary number
 of degrees of freedom, but a clear geometrical interpretation of this
 integrability-preserving deformation procedure was still lacking.

 The aim of this letter is to show a neat connection between
 two-dimensional (2D)
 spaces with
 non-constant curvature and Poisson coalgebra deformations. In
 particular, we will
 show that a certain class of  $q$-Poisson coalgebras $(q={\rm e}^z)$
 generates in a very natural way a family of integrable geodesic
 motions on 2D manifolds with a (in general,
 non-constant) curvature $K$ that turns out to be a function of the
 deformation parameter $z$. Moreover, such a curvature is directly
 generated by the ``twisted" coproduct map of the deformed coalgebra.
 We also stress that, as a consequence of coalgebra symmetry, a
 straightforward generalization of this construction to arbitrary
 dimensions can be obtained, whose complete description will be
 presented elsewhere~\cite{ND}.

 Let us briefly recall  the basics of the construction of Hamiltonian
 systems with coalgebra symmetry by using the  non-deformed Poisson
 coalgebra
 $(sl(2), \Delta)$, which is  defined by the following Poisson
 brackets and coproduct map
 $\Delta$:
 \bea
 &&\{\jj,\jp\}=2\jp  ,\qquad 
 \{\jj,\jm\}=-2\jm  ,\qquad \{\jm,\jp\}=4\jj  ,\label{aa} \\ 
 &&\co(J_i)= J_i\otimes 1+1\otimes J_i  ,\qquad i=+,-,3 .
 \label{ab}
 \eea
 The  Casimir function for this Poisson coalgebra is ${\cal C}= \jm
 \jp-\jj^2 $. A one-particle  symplectic realization  of (\ref{aa}) is
 given by 
 \be
  \jm^{(1)}=q_1^2 , \qquad
 \jp^{(1)}=p_1^2 +
 \frac{\otra_1}{q_1^2} , \qquad
 \jj^{(1)}= q_1 p_1   ,
 \label{ac}
 \ee
 where  $\otra_1$ is just the constant that labels the phase space
 realization:
 ${\cal C}^{(1)}=\otra_1$.   The corresponding two-particle
 realization  is obtained through the coproduct (\ref{ab})   by using
 one   symplectic realization for each lattice site: 
 \be
 \jm^{(2)}= 
 q_1^2 + q_2^2 ,  \qquad
 \jp^{(2)} = 
 p_1^2 + p_2^2 + \frac{\otra_1}{q_1^2}+ \frac{\otra_2}{q_2^2} ,
 \qquad  \jj^{(2)}= q_1 p_1 + q_2 p_2 .
 \label{ad}
 \ee
 Given any Hamiltonian function ${\cal H}$ on the generators of
 $(sl(2), \Delta)$, the coalgebra symmetry ensures that the associated
 two-body Hamiltonian
 ${\cal H}^{(2)}:=
 \Delta({\cal H})={\cal
 H}(\jm^{(2)},\jp^{(2)},\jj^{(2)})$
  is integrable  since the two-particle Casimir
 \be  
  {\cal C}^{(2)}= \Delta({\cal C}) =({q_1}{p_2} - {q_2}{p_1})^2
 + \left(
 \otra_1\frac{q_2^2}{q_1^2}+\otra_2\frac{q_1^2}{q_2^2}\right)
 + \otra_1+ \otra_2  ,
 \label{af}
 \ee
 Poisson commutes with ${\cal H}^{(2)}$ with respect to the bracket
 $
 \{f,g\}=\sum_{i=1}^2\left(\frac{\partial f}{\partial q_i}
 \frac{\partial g}{\partial p_i}
 -\frac{\partial g}{\partial q_i} 
 \frac{\partial f}{\partial p_i}\right).
 $

 Some well-known (super)integrable Hamiltonian systems can
 be recovered as specific choices for
 ${\cal H}^{(2)}$. In particular, if we set 
 \be
 {\cal H}=\frac 12 \jp  +{\cal F}\left(\jm  \right),
 \label{ahh}
 \ee
 where ${\cal F}$ is an arbitrary smooth function, we find the
 following family of integrable systems defined on the  2D  Euclidean
 space  $\>E^2$:
 \be
 {\cal H}^{(2)}= \frac 12 \left( p_1^2 + p_2^2 \right) +
 \frac{\otra_1}{2 q_1^2}+\frac{\otra_2}{2 q_2^2} +{\cal
 F}\left(q_1^2+q_2^2 \right) .
 \label{ai}
 \ee
 The case ${\cal F}\left(\jm  \right)=\omega^2 \jm=\omega^2
 (q_1^2+q_2^2)$ gives rise to the 2D Smorodinsky--Winternitz  
 system~\cite{Fris,Evansa,Evansb}. Obviously, the free motion on $\>E^2$  is
 described by ${\cal H}=\frac 12 \jp$. The $N$-body generalization of this
 construction follows from   iteration of the coproduct map~\cite{Deform}.

 The very same coalgebra approach~\cite{BR} leads to integrable
 deformations of (\ref{ai}) by considering deformations of $sl(2)$
 coalgebras. This was the procedure applied in~\cite{Deform} through
 the Poisson analogue $(sl_z(2),
 \Delta_z)$ of the non-standard quantum deformation of
 $sl(2)$~\cite{Ohn}:
 \be 
 \{\jj,\jp\}=2 \jp \cosh z\jm , \qquad 
  \{\jj,\jm\}=-2\,\frac {\sinh z\jm}{z} ,\qquad
 \{\jm,\jp\}=4 \jj   ,
 \label{ba}
 \ee 
 \be  
  \Delta_z(\jm)=  \jm \otimes 1+
 1\otimes \jm  ,\qquad 
 \Delta_z(J_i)=J_i \otimes {\rm e}^{z \jm} + {\rm e}^{-z \jm} \otimes
 J_i , \quad\ i=+,3  ,
 \label{bb}
 \ee 
 where $z=\ln q$ is the deformation parameter. Hereafter  we shall
 assume  that  
 $z\in \R$. The deformed  Casimir function reads
 \be
 {\cal C}_z= \frac {\sinh z\jm}{z}\, \jp -\jj^2  .
 \label{bc}
 \ee
 Since in the following we shall consider only free motion, we
 restrict to the
 $\otra_i=0$ case. Then, a one-particle symplectic realization of
 $sl_z(2)$ turns out to be~\cite{Deform}
 \be  
 \jm^{(1)}=q_1^2 ,\qquad
 \jp^{(1)}=\frac {\sinh z q_1^2}{z q_1^2}\, p_1^2 ,\qquad
 \jj^{(1)}= \frac {\sinh z q_1^2}{z q_1^2}\, q_1 p_1  ,
 \label{bd}
 \ee 
 where  ${\cal C}_z^{(1)}=0$. Hence dimensions of  $z$  are
 $[z]=[q_1]^{-2}=[\jm]^{-1}$. Next,  the deformed coproduct $\Delta_z$
 provides  the following two-particle symplectic realization of
 (\ref{ba}):
 \be  
 \begin{array}{l}
 \displaystyle{ \jm^{(2)}=q_1^2+q_2^2 ,\qquad  \jp^{(2)}=
  \frac {\sinh z q_1^2}{z q_1^2}\,  {\rm e}^{z q_2^2}   p_1^2   +
 \frac {\sinh z q_2^2}{z q_2^2} \,  {\rm e}^{-z q_1^2}   p_2^2  
 },\\[10pt]
 \displaystyle{\jj^{(2)}=
 \frac {\sinh z q_1^2}{z q_1^2 } \, {\rm e}^{z q_2^2}   q_1 p_1  +
 \frac {\sinh z q_2^2}{z q_2^2 }\, {\rm e}^{-z q_1^2}    q_2 p_2  }.
 \end{array}
 \label{be}
 \ee 
 Consequently, the two-particle Casimir given by  
 \bea
 && {\cal C}^{(2)}_z =  \Delta_z({\cal C}_z)= \frac {\sinh z q_1^2
 }{z q_1^2 } \,
 \frac {\sinh z q_2^2}{z q_2^2}\,{\rm e}^{-z q_1^2}{\rm e}^{z q_2^2}  
 \left({q_1}{p_2} - {q_2}{p_1}\right)^2 ,
 \label{bf}
 \eea
 is,  by construction, a constant of the motion for any Hamiltonian
 ${\cal H}^{(2)}_z=\Delta_z({\cal H})={\cal
 H}(\jm^{(2)},\jp^{(2)},\jj^{(2)})$. 

 Thus, by taking into account the explicit expressions (\ref{be}), we find
 that   the most general   {\em integrable} (and quadratic
 in the momenta) deformation  of  the free motion on
 $\>E^2$ with $(sl_z(2),
 \Delta_z)$-symmetry reads 
 \be
 {\cal H} =\frac 12 \jp f(z\jm),
 \label{bbff}
 \ee
 where $f$ is any smooth function such that 
 $\lim_{z\to 0}f(z\jm)=1$ (note that $\lim_{z\to 0}\jp=p_1^2+p_2^2$). The
 simplest choice of (\ref{bbff}) corresponds to taking ${\cal H}=\frac 12
 \jp$, namely 
 \be
 {\cal H}^{\rm \SW}_z =\frac12 \left( \frac {\sinh z
 q_1^2}{z q_1^2} \, {\rm e}^{z q_2^2}  p_1^2   +
 \frac {\sinh z q_2^2}{z q_2^2} \, {\rm e}^{-z q_1^2}  p_2^2  \right)   .
 \label{bg}
 \ee 
 On the other hand, a further analysis of (\ref{bbff}) leads to a
 {\em superintegrable} deformation of  the free
 Euclidean motion which is provided by
 ${\cal H}=\frac 12  \jp {\rm e}^{ z \jm} $, that is,
 \be
  {\cal H}^{\rm \Stc}_z =\frac12 \left( \frac {\sinh z
 q_1^2}{z q_1^2} \, {\rm e}^{z q_1^2}{\rm e}^{2z q_2^2}  p_1^2   +
 \frac {\sinh z q_2^2}{z q_2^2} \, {\rm e}^{ z q_2^2}  p_2^2  \right) . 
 \label{bi}
 \ee 
 In this case, besides (\ref{bf}), there exists  the additional
 constant of the motion~\cite{Deform}:
 \be
 {\cal I}_z=\frac {\sinh z
 q_1^2}{2 z q_1^2} \, {\rm e}^{z q_1^2}  p_1^2 .
 \label{bjj}
 \ee
 As ${\cal C}^{(2)}_z$, ${\cal I}_z$ and ${\cal H}^{\rm \Stc}_z$ are
 functionally independent functions, the latter is a (maximally)
 superintegrable Hamiltonian.

 It becomes clear that (\ref{bbff}) and, consequently, both Hamiltonians 
 ${\cal H}^{\rm
 \SW}_z $ and ${\cal H}^{\rm \Stc}_z$, can be interpreted as a deformed kinetic
 energy ${\cal T}_z(q_i,p_i)$  in such a way that 
 a  ``deformation" of $\>E^2$ arises from the dynamics.
 In fact,
 ${\cal T}_z(q_i,p_i)\to {\cal T}(p_i)=\frac 12 (p_1^2 + p_2^2)$ under
 the limit
 $z\to 0$. In the sequel we shall unveil the
 ``hidden" supporting spaces related to  ${\cal T}_z$ coming from
 (\ref{bg}) and
 (\ref{bi}).  In particular, in Section 2  we shall show  that the
 Hamiltonian
 ${\cal H}^{\rm
 \SW}_z $ will give  rise  to integrable geodesic motions on 2D
 Riemannian spaces and $(1+1)$D relativistic spacetimes, all of them
 with a non-constant curvature governed by  the   deformation parameter
 $z$. Section 3 will be devoted to the Hamiltonian ${\cal H}^{\rm
 \Stc}_z$, which  will   provide   superintegrable geodesics on the 
 sphere and hyperbolic spaces as well as on the   (anti)de Sitter and
 Minkowskian spacetimes with a curvature exactly equal to  $z$. Some comments
 concerning   other  possible particular Hamiltonians contained in the
 family (\ref{bbff}) (including the general expression for the
 associated curvature) and   several open problems close the letter.

 \sect{Integrable deformation and   non-constant curvature}

 The kinetic energy ${\cal T}^{\rm \SW}_z(q_i,p_i)$ coming from
 (\ref{bg})  can be rewritten as the free Lagrangian
 \be
 {\cal T}^{\rm \SW}_z(q_i,\dot q_i)=\frac 12 \left(\frac
 {z q_1^2}{\sinh z q_1^2} \, {\rm e}^{-z q_2^2} \dot q_1^2   +
 \frac {z q_2^2}{\sinh z q_2^2} \, {\rm e}^{z q_1^2} \dot q_2^2  
 \right) ,
 \label{ca}
 \ee 
 that defines a geodesic flow on a 2D
 Riemannian space with a definite positive  metric  with signature  
 diag$(+,+)$ given, up to a constant factor, by
 \be
 \d s^2=\frac {2z q_1^2}{\sinh z
 q_1^2} \, {\rm e}^{-z q_2^2} \,\d q_1^2   +
  \frac {2 z q_2^2}{\sinh z q_2^2} \, {\rm e}^{z q_1^2}\, \d q_2^2  .
 \label{cc}
 \ee
 If we write the metric as $ \d s^2=g_{11}(q_1,q_2)\d
 q_1^2+g_{22}(q_1,q_2)\d q_2^2$, the Gaussian curvature $K$ can be
 directly computed by using the formula~\cite{Berry}
 \be
 K=\frac{-1}{\sqrt{g_{11} g_{22}}}\left\{ \frac{\partial}{\partial
 q_1} \left( \frac{1}{\sqrt{g_{11}}} \frac{\partial
 \sqrt{g_{22}}}{\partial q_1}
 \right)+
 \frac{\partial}{\partial
 q_2} \left( \frac{1}{\sqrt{g_{22}}} \frac{\partial
 \sqrt{g_{11}}}{\partial q_2}
 \right)\right\} ,
 \label{ccc}
 \ee
 which gives a non-constant and {\em negative} curvature
  \be
 K(q_1,q_2;z)=-   z \sinh\left(z(q_1^2+q_2^2) \right) .
 \label{cd}
 \ee
 Thus the 
 underlying 2D space is of hyperbolic  type and  
 endowed with  a radial symmetry. We stress that the exponentials in 
 the metric (\ref{cc}) are essential in order to obtain a non-vanishing
 curvature $K$. Such exponentials are indeed the ones appearing in the
 deformed coproduct (\ref{bb}), thus showing the direct connection
 between coproduct deformation and curved spaces.

 Let us introduce a change of coordinates that incorporate, besides
 $z$, another parameter $\la_2\ne 0$. In particular, we write
 $z=\la_1^2$ and   consider a pair of  new coordinates $(\rr,\te)$
 defined through the expressions
 \be
 \cosh(\la_1 \rr)=\exp\left\{z(q_1^2+q_2^2)\right\} ,\qquad
 \sin^2(\la_2 \te)=\frac{\exp\left\{2z q_1^2
 \right\}-1}{\exp\left\{2z(q_1^2+q_2^2)\right\}-1},
 \label{ce}
 \ee
 where both $\la_1=\sqrt{z}$ and $\la_2$ can take either a real or a
 pure imaginary value. In this way, we will be able to rewrite the
 initial metric (\ref{cc}) as a family of six metrics on spaces with
 different signature and curvature.  The geometrical meaning  of
 $(\rr,\te)$ can be  appreciated by taking the first-order terms of the
 expansion in $z$ of (\ref{ce}):
 \be
 \rr^2\simeq 2(q_1^2+q_2^2),\qquad \sin^2(\la_2
 \te)\simeq\frac{q_1^2}{q_1^2+q_2^2}.
 \label{cf}
 \ee
 Therefore $\rr$ can be interpreted as a radial coordinate and $\te$
 is a  circular or  hyperbolic angle for either a real or an imaginary
 $\la_2$, respectively. At this first-order level, the  ``Cartesian"
 coordinates would be
 $(x,y)=\sqrt{2}(q_2,q_1)$.

 Under the
 transformation (\ref{ce}),  the metric (\ref{cc}) takes a simpler
 form:
 \be
 \d s^2=\frac {1}{\cosh(\la_1 \rr)}
 \left( \d \rr^2  +\la_2^2\,\frac{\sinh^2(\la_1 \rr)}{\la_1^2} \, \d
 \te^2  \right) =\frac {1}{\cosh(\la_1 \rr)}\,\d s_0^2.
 \label{cg}
 \ee
 Now, if we recall the description of the 2D Cayley--Klein (CK) spaces
 in terms of geodesic polar coordinates~\cite{Trigo,Conf} (these
 spaces are parame\-trized by their constant curvature $\kappa_1$ and a
 second real parameter
 $\kappa_2$), we realize that $\d
 s_0^2$ is just the metric of the CK spaces
 provided that we identify $z=\la_1^2\equiv -\kappa_1$ and
 $\la_2^2\equiv \kappa_2$. In particular, from (\ref{cg}) and by taking
 into account the admissible specializations of
 $z$ and $\lambda_2$, we find the following underlying spaces:

 \begin{itemize}

 \item[$\bullet$]  When  $\la_2$ is real, we get a  2D deformed
 sphere ${\bf S}^2_z$ $(z<0)$,    and  a deformed
 hyperbolic or  Lobachewski space ${\bf H}^2_z$ $(z>0)$.

 \item[$\bullet$]  When  $\la_2$ is imaginary, we obtain a deformation
 of   the (1+1)D   anti-de Sitter spacetime ${\bf
 AdS}_z^{1+1}$ $(z<0)$ and of the   de Sitter one ${\bf dS}_z^{1+1}$
 $(z>0)$.

 \item[$\bullet$] In the non-deformed case $z\to 0$,  the Euclidean
 space
 ${\bf E}^2$ 
 ($\la_2$ real) and  Minkowskian spacetime
 ${\bf M}^{1+1}$ ($\la_2$ imaginary) are recovered.

 \end{itemize}
  
 Thus the ``additional" parameter $\la_2$  governs the
 signature of the metric. The case $\la_2=0$, that we do not consider
 here, corresponds to Newtonian spacetimes endowed with a degenerate
 metric (see (\ref{cg})). 

 In the new coordinates, the sectional (Gaussian) curvature reads
 \be
 K(\rr)=-\frac 12 \la_1^2 \,\frac{\sinh^2(\la_1 \rr)}{\cosh(\la_1
 \rr)} ,
 \label{cj}
 \ee
 and the scalar curvature is just $2K(\rr)$. Hence the behaviour of
 the function
 $K(\rr)$ has a non-trivial dependence on   the sign of the
 deformation parameter:

 \begin{itemize}

 \item[$\bullet$]  If $z$ is positive then $\la_1$ is a real number
 and $K(\rr)$ is always  an increasing {\em negative} function that
 goes from $K=0$ at the origin
 $\rr=0$ up to $K\to -\infty$ when $\rr\to +\infty$.

 \item[$\bullet$] If $z$ is negative then $\la_1$ is a pure imaginary
 number, and
 $K(\rr)$ is a periodic function with   single poles at the points
 $|\la_1| \rr=\frac{\pi}{2}, \frac{3\pi}{2},
 \dots$, and (double) zeros at   $|\la_1| \rr=0, \pi , 2\pi , \dots$. 
 Then
 $K(\rr)$ is   
  {\em negative}   in the intervals $|\la_1|\rr\in(0, \frac{\pi}{2}
 ),(\frac{3\pi}{2}, \frac{5\pi}{2}),\dots$ but it is 
  {\em positive}   when $|\la_1|\rr\in  (\frac{\pi}{2},
 \frac{3\pi}{2}), (\frac{5\pi}{2}, \frac{7\pi}{2}),\dots$ 

  \end{itemize}

 In Table \ref{table1} we display the metric (\ref{cg}) and the
 sectional curvature (\ref{cj}) for  the six particular spaces that
 arise according to 
 $(\la_1,\la_2)$ by considering the simplest values  $\la_i\in\{1,{\rm
 i}\}$.
  In the deformed
 Riemannian    spaces ($\la_2=1$) the metric is   always a positive
 definite one on ${\bf H}_z^2$, while on 
 ${\bf S}_z^2$ this can be either a positive or negative definite
 metric in the  intervals with
 $\rr\in(0, \frac{\pi}{2}
 ),(\frac{3\pi}{2}, \frac{5\pi}{2}),\dots$ and $\rr\in 
 (\frac{\pi}{2}, \frac{3\pi}{2}), (\frac{5\pi}{2},
 \frac{7\pi}{2}),\dots$    respectively. Likewise, in the deformed  
 spacetimes ($\la_2={\rm i}$) we obtain  a Lorentzian metric  that
 keeps its global sign on
 ${\bf dS}_z^{1+1}$ but alternates it on ${\bf AdS}_z^{1+1}$ in the
 same previous intervals.
      The {\em contraction} $\la_1\to 0$
 ($z\to 0$) in each column of table \ref{table1} gives either the
 proper Euclidean or the Minkowskian space as the  limiting
 non-deformed/flat case; in the latter the ``angle" $\te$ is indeed a
 rapidity
  in units $c=1$.

 \begin{table}[t]
 {\footnotesize
  \noindent
 \caption{{Metric and sectional curvature of the underlying spaces for
 different values of the deformation parameter $z=\lambda_1^2$ and
 signature parameter $\lambda_2$.}}
 \label{table1}
 \medskip
 \noindent\hfill
 $$
 \begin{array}{ll}
 \hline
 \\[-6pt]
 {\mbox {2D deformed Riemannian spaces}}&\quad{\mbox  {$(1+1)$D
 deformed relativistic spacetimes}}\\
 [4pt] 
 \hline
 \\
 [-6pt]
 \mbox {$\bullet$ Deformed sphere ${\bf S}^2_z$}&\quad\mbox {$\bullet$
 Deformed anti-de Sitter spacetime ${\bf AdS}^{1+1}_z$}\\[4pt] z=-1;\
 (\la_1,\la_2)=({\rm i},1)&\quad z=-1;\ (\la_1,\la_2)=({\rm i},{\rm
 i})\\[4pt]
 \displaystyle{\d s^2 =\frac{1}{\cos \rr}\left( \d \rr^2+\sin^2
 \rr\,\d\te^2
 \right)} &\quad
 \displaystyle{\d s^2 =\frac{1}{\cos \rr}\left( \d \rr^2-\sin^2
 \rr\,\d\te^2
 \right)} \\[8pt]
  \displaystyle{K =-\frac{\sin^2 \rr}{2\cos \rr} } &\quad
  \displaystyle{K =-\frac{\sin^2 \rr}{2\cos \rr} } \\[12pt]
 \mbox {$\bullet$ Euclidean space  ${\bf E}^2$}&\quad\mbox {$\bullet$
 Minkowskian spacetime ${\bf M}^{1+1}$}\\[4pt]
 z=0;\ (\la_1,\la_2)=(0,1)&\quad
 z=0;\ (\la_1,\la_2)=(0,{\rm i})\\[4pt]
  \displaystyle{\d s^2 =  \d \rr^2+ \rr^2\d\te^2
  } &\quad
  \displaystyle{\d s^2 =  \d \rr^2- \rr^2\d\te^2} \\[2pt]
  \displaystyle{K =0 } &\quad
  \displaystyle{K =0} \\[6pt]
 \mbox {$\bullet$ Deformed hyperbolic space ${\bf H}_z^2$}&\quad\mbox
 {$\bullet$ Deformed de Sitter spacetime ${\bf dS}^{1+1}_z$}\\[4pt]
 z=1;\ (\la_1,\la_2)=(1,1)&\quad
 z=1;\ (\la_1,\la_2)=(1,{\rm i})\\[4pt]
 \displaystyle{\d s^2 =\frac{1}{\cosh \rr}\left( \d \rr^2+\sinh^2
 \rr\,\d\te^2
 \right)} &\quad
 \displaystyle{\d s^2 =\frac{1}{\cosh \rr}\left( \d \rr^2-\sinh^2
 \rr\,\d\te^2
 \right)} \\[8pt]
 \displaystyle{K =-\frac{\sinh^2 \rr}{2\cosh \rr} } &\quad
  \displaystyle{K =-\frac{\sinh^2 \rr}{2\cosh \rr} } \\[8pt]
 \hline
 \end{array}
 $$
 \hfill}
 \end{table}

 The metric (\ref{cg}) gives rise to the kinetic term of ${\cal
 H}^{\SW}_z$ in the new coordinates $(\rr,\te)$:
 \be
 {\cal T}^{\rm \SW}_z(\rr,\te;\dot \rr,\dot \te)=\frac
 {1}{2\cosh(\la_1 \rr)}
 \left(\dot \rr^2  +\la_2^2\,\frac{\sinh^2(\la_1 \rr)}{\la_1^2} \,
 \dot \te^2 
 \right) .
 \label{da}
 \ee
 It is a matter of straightforward computation to obtain the new
 Hamiltonian, that we define as  $\widetilde{H}^{\SW}_z(\rr,\te;p_\rr,
 p_\te)=2 {\cal H}^{\SW}_z(q_i,p_i)$; this reads
 \be 
 \widetilde{H}^{\SW}_z=\frac 12 \cosh(\la_1
 \rr)\left(p_\rr^2 +\frac{\la_1^2}{\la_2^2\sinh^2(\la_1 \rr)} \, 
 p_\te^2\right).
 \label{dd}
 \ee
 The corresponding constant of the motion comes from (\ref{bf}). If we  
 denote the new integral as
  $ \widetilde{C}_z(\rr,\te;p_\rr, p_\te)=4\la_2^2  {\cal
 C}^{(2)}_z(q_i,p_i)$  it can be shown that $\widetilde{C}_z=p_\te^2 $
 which, in turn, allows us to perform the usual reduction of (\ref{dd})
 to the 1D (radial) Hamiltonian given by
 \be 
 \widetilde{H}^{\SW}_z =\frac 12 \cosh(\la_1
 \rr)\, p_\rr^2  +\frac{\la_1^2 \cosh(\la_1
 \rr)}{2\la_2^2\sinh^2(\la_1 \rr)} \, \widetilde{C}_z.
 \label{dede}
 \ee 
 The integration of the geodesic motion on all these spaces can be
 explicitly performed in terms of elliptic integrals, and it will be
 fully described elsewhere~\cite{ND}.

 \sect{Superintegrable deformation and constant curvature}

 Let us consider now the superintegrable Hamiltonian (\ref{bi}). The
 free Lagrangian ${\cal T}^{\rm \Stc}_z$  turns out to be
 \be
 {\cal T}^{\rm \Stc}_z(q_i,\dot q_i)=\frac 12 \left(\frac {z
 q_1^2}{\sinh z q_1^2} \, {\rm e}^{-z q_1^2}{\rm e}^{-2 z q_2^2} \,\dot
 q_1^2   +
 \frac {z q_2^2}{\sinh z q_2^2} \, {\rm e}^{-z q_2^2}\, \dot q_2^2  
 \right)  .
 \label{ea}
 \ee 
   Thus the associated metric is given by
 \be
 \d s^2=\frac {2z q_1^2}{\sinh z
 q_1^2} \, {\rm e}^{-z q_1^2}{\rm e}^{-2 z q_2^2} \,\d q_1^2   +
  \frac {2 z q_2^2}{\sinh z q_2^2} \, {\rm e}^{-z q_2^2} \, \d q_2^2  
 .
 \label{ec}
 \ee
 Remarkably enough, in this case the    Gaussian curvature (obtained
 by applying (\ref{ccc})) is {\em constant} and coincides with the
 deformation parameter $K=z$.
  
 Under the change of coordinates (\ref{ce}), $(q_1,q_2)\to (\rr,\te)$,
 the metric (\ref{ec}) becomes
 \be
 \d s^2=\frac {1}{\cosh^2(\la_1 \rr)}
 \left( \d \rr^2  +\la_2^2\,\frac{\sinh^2(\la_1 \rr)}{\la_1^2} \, \d
 \te^2  \right)  =\frac {1}{\cosh^2(\la_1 \rr)}\,\d s_0^2 ,
 \label{ee}
 \ee
 where $\d s_0^2$ is again the metric of the 2D  CK 
 spaces. As these spaces are also of constant curvature, a further
 change of coordinates  should allow us to reproduce exactly the CK
 metric.  This can be achieved by introducing a new radial coordinate
 $r$ as
 \be
 r=\int_0^{\rr}\frac{\d x}{\cosh(\la_1 x)}  ,
 \label{ef}
 \ee
 which for $\la_1=1$ is the Gudermannian function, while for 
 $\la_1={\rm i}$ is the   lambda function~\cite{Conf,Dwight}. By
 making use of the  functional relations 
 \be
 \tanh\left(\la_1\frac{\rr}{2}\right)
 =\tan\left(\la_1\frac{r}{2}\right),
 \quad\ 
  \cosh(\la_1 \rr ) =\frac{1}{\cos(\la_1 r)} ,\quad\ 
 \sinh(\la_1 \rr )=\tan(\la_1 r) ,
 \label{eg}
 \ee
 we finally obtain
 \be
 \d s^2= 
  \d r^2  +\la_2^2\,\frac{\sin^2(\la_1 r)}{\la_1^2} \, \d \te^2   ,
 \label{eh}
 \ee
 which is just the CK metric written in  geodesic polar coordinates
 $(r,\te)$ and provided that  $z=\la_1^2\equiv \kappa_1$ and
 $\la_2^2\equiv 
 \kappa_2$~\cite{Conf}. Note that in this case  $z=\kappa_1$,
 in contrast with   the previous Section where
 $z=-\kappa_1$; this is due to the interchange between  circular and
 hyperbolic trigonometric functions  (see (\ref{eg})) entailed by the
 definition (\ref{ef}). Notice also that in the limiting case    $z\to
 0$ the coordinate  $\rr\to r$.

  The kinetic energy in the new coordinates  reads
 \be
 {\cal T}^{\rm \Stc}_z(r,\te;\dot r,\dot \te)=\frac {1}{2}
 \left(\dot r^2  +\la_2^2\,\frac{\sin^2(\la_1 r)}{\la_1^2} \, \dot
 \te^2 
 \right)  .
 \label{fa}
 \ee
 The  Hamiltonian ${\cal H}^{\rm \Stc}_z$  (\ref{bi}) and
 its constants of the motion   ${\cal C}^{(2)}_z$ (\ref{bf}) and 
 ${\cal I}_z$ (\ref{bjj}) are expressed in canonical geodesic polar
 coordinates
 $(r,\te)$ and momenta $(p_r,p_\te)$ as 
 \be  
 \begin{array}{l}
 \displaystyle{
 \widetilde{H}^{\Stc}_z=\frac 12 \left(p_r^2
 +\frac{\la_1^2}{\la_2^2\sin^2(\la_1 r)}
 \,  p_\te^2\right) ,\qquad  \widetilde{C}_z=p_\te^2 , }\\[10pt]
 \displaystyle{ 
  \widetilde{I}_z=\left(\la_2\sin(\la_2\te)
 p_r+\frac{\la_1\cos(\la_2\te)}{\tan(\la_1 r)}\,p_\te
 \right)^2 ,}
 \end{array}
 \label{fn}
 \ee 
  where we have denoted
 $\widetilde{H}^{\Stc}_z=2{\cal H}^{\Stc}_z$, 
 $\widetilde{C}_z=4\la_2^2 {\cal C}^{(2)}_z$ and
 $\widetilde{I}_z=4\la_2^2 {\cal I}_z  $.
 Then 
 $\widetilde{H}^{\Stc}_z$ is transformed into a ``radial" 1D system
 from which the geodesic curves can   be    obtained~\cite{ND}. Namely,
 \be 
 \widetilde{H}^{\Stc}_z=\frac 12 \, p_r^2
 +\frac{\la_1^2}{2\la_2^2\sin^2(\la_1 r)} \, 
  \widetilde{C}_z .
 \label{ffnn}
 \ee

 \sect{Concluding remarks}

 Some short comments and remarks are in order. 
 Firstly,  we stress that have worked out the geometric interpretation
 of two outstanding representatives among the class of
 Hamiltonians (\ref{bbff}). However,   the general expression (\ref{bbff})
 comprises   many other possible choices for a deformed kinetic energy and
 therefore for the ``dynamical" generation of deformed spaces.   A
 preliminary analysis can be performed   by considering the
 general expression of the   curvature coming from (\ref{bbff}), which
 turns out to be  
 \be
 K(x)=\frac{z}{f(x)}\left( f(x)f^\prime(x)\cosh x  +\left(
  f(x)f^{\prime\prime}(x)-f^2(x)-{f^\prime}^2(x)
 \right) \sinh x
 \right) ,
 \ee
 where $x\equiv z\jm=z(q_1^2+q_2^2)=z q^2$, $f^\prime=\frac{{\rm
 d}f(x)}{{\rm d}
 x}$ and $f^{\prime\prime}=\frac{{\rm d}^2f(x)}{{\rm d} x^2}$.
 Now  if we take,
 for instance,
 ${\cal H}=\frac 12  \jp {\rm e}^{ z \alpha \jm} $,
  where $\alpha$ is an extra real parameter, we obtain the following 
 Gaussian curvature
 \be
 K(q^2;z)=z\,{\rm e}^{z \alpha q^2}\left( \alpha \cosh(z q^2)-\sinh(z
 q^2)\right),
 \ee
 with   power series
 expansion in $z$ given by
  \be
 K(q^2;z)=\alpha\, z + (\alpha^2 - 1) q^2\,z^2 + \frac \alpha 2 (\alpha^2 -
 1) q^4\,z^3 + o[z^4] .
 \label{cdb}
 \ee
 Thus  the cases $\alpha=\pm 1$ are the only ones with constant
 $K$ ($\alpha= 1$ is just ${\cal H}^{\rm \Stc}_z$) and $\alpha\neq \pm 1$
 defines a class of spaces with non-constant curvature that includes 
 ${\cal H}^{\rm \SW}_z $ for $\alpha=0$. Another example is provided by
 ${\cal H}=\frac 12  \jp \cosh({ z \beta \jm}) $ where $\beta$ can be
 either a real or a  pure imaginary   parameter. The curvature and   its power
 series expansion read  
 \be  
 \begin{array}{l}
 \displaystyle{K(q^2;z)=z\left(   \beta^2- \cosh^2(z \beta q^2)     
 \right)\frac{\sinh(zq^2)}{\cosh(z \beta q^2)}+z\beta\cosh(z 
 q^2)\sinh(z\beta q^2)}\\[12pt]
 \qquad\quad\ \ \, =(2 \beta^2-1) q^2\,z^2 - \frac 1 6 (2 \beta^4
 -\beta^2+1)
  q^6\,z^4 + o[z^6] ,
 \end{array}
 \ee 
 so that  the underlying deformed spaces   are always of non-constant
 curvature  (for any  $\beta$).

 We also stress that the generalization to arbitrary dimensions can
 be   readily obtained from the $N$-degrees of freedom symplectic
 realization of the $(sl_z(2),
 \Delta_z)$ coalgebra given in~\cite{Deform}. For example, the
 explicit form of the
 $N$-body kinetic energy arising from the Hamiltonian
 ${\cal H}=\frac 12 
 \jp$ would be:
 $$
 {\cal H}^{{\rm
 \SW},(N)}_z = \frac 12 
 \sum_{i=1}^N
  \frac {\sinh z q_i^2}{z q_i^2}\,  p_i^2    \exp{\left( - z
 \sum_{k=1}^{i-1}  q_k^2+  z \sum_{l=i+1}^N   q_l^2 \right)} .
 \label{ddr}
 $$
 The exponentials coming from the coproduct are the 
 objects that generate the non-vanishing sectional curvatures.  The
 geometric characterization of the underlying $N$-dimensional curved
 spaces is under investigation~\cite{ND}.

 On the other hand,   by considering the deformation of the more
 general symplectic realization (\ref{ac}) with $b_i\neq
 0$~\cite{Deform}, together with a Hamiltonian of the type
 (\ref{ahh}),  one could obtain the non-constant curvature analogues of
 the Smorodinsky--Winternitz potential. We also mention that the study
 of the free motion of a quantum mechanical particle on the curved
 spaces here introduced should provide a geometric interpretation of
 the quantum analogue of the Poisson algebra (\ref{ba}), which is just
 the non-standard quantum deformation of
 $sl(2)$~\cite{Ohn}. Work on these lines is in progress.

 Finally, one could consider the
 Poisson algebra (\ref{ba}) as a Poisson--Lie structure on a dual
 group $G^\ast$ with Lie algebra given by the dual of the Lie bialgebra
 cocommutator associated to the coproduct (\ref{bb})~\cite{PL}. In
 other words, (\ref{ba}) is a sub-Poisson coalgebra of the full
 canonical Poisson--Lie structure on the Drinfel«d double
 $D_{ns}(su(2))$ associated to the non-standard quantum deformation of
 $sl(2)$ (as a real Lie group, $D_{ns}(su(2))$ was proven to be
 isomorphic  to a (2+1)D Poincar\'e group~\cite{gomez}). Since
 $\sigma$-models related by Poisson--Lie T-duality are directly
 connected to  canonical Poisson--Lie structures on Drinfel'd
 doubles~\cite{duality,Lledo,sfetsos,Snobl}, the construction of the
 $\sigma$-model associated to
 $D_{ns}(su(2))$ and its relationship with the results here presented
 could be worth to be considered.

 {\small

 \section*{Acknowledgements}

 This work was partially supported  by the Ministerio de
 Educaci\'on y Ciencia   (Spain, Project FIS2004-07913),  by the Junta
 de Castilla y Le\'on   (Spain, Project  BU04/03), and by INFN-CICyT
 (Italy-Spain).

 }



 \end{document}